\documentclass[preprint,prc,aps,showpacs,groupedaddress,superscriptaddress,floatfix]{revtex4}

\input{tcilatex}

\begin{document}

\title{Approximated $l$-states of the Manning-Rosen potential by
Nikiforov-Uvarov method}
\author{Sameer M. Ikhdair}
\email[E-mail: ]{sikhdair@neu.edu.tr}
\affiliation{Department of Physics, Near East University, Nicosia, North Cyprus, Turkey}
\date{\today}

\begin{abstract}
The approximately analytical bound state solutions of the $l$-wave Schr\"{o}%
dinger equation for the Manning-Rosen (MR) potential are carried out by a
proper approximation to the centrifugal term. The energy spectrum formula
and normalized wave functions expressed in terms of the Jacobi polynomials
are both obtained for the application of the Nikiforov-Uvarov (NU) method to
the Manning-Rosen potential. To show the accuracy of our results, we
calculate the eigenvalues numerically for arbitrary quantum numbers $n$ and $%
l$ with two different values of the potential parameter $\alpha .$ It is
found that our results are in good agreement with the those obtained by
other methods for short potential range, small $l$ and $\alpha .$ Two
special cases are investigated like the $s$-wave case and Hulth\'{e}n
potential case.

\end{abstract}

\maketitle

\bigskip

\section{Introduction}

\noindent One of the important tasks of quantum mechanics is to find exact
solutions of the wave equations (nonrelativistic and relativistic) for
certain type of potentials of physical interest since they contain all the
necessary information regarding the quantum system under consideration. For
example, the exact solutions of these wave equations are only possible in a
few simple cases such as the Coulomb, the harmonic oscillator,
pseudoharmonic and Mie-type potentials [1-8]. For an arbitrary $l$-state,
most quantum systems could be only treated by approximation methods. For the
rotating Morse potential some semiclassical and/or numerical solutions have
been obtained by using Pekeris approximation [9-13]. In recent years, many
authors have studied the nonrelativistic and relativistic wave equations
with certain potentials for the $s$- and $l$-waves. The exact and
approximate solutions of these models have been obtained analytically
[10-14].

Many exponential-type potentials have been solved like the Morse potential
[12,13,15], the Hulth\'{e}n potential [16-19], the P\"{o}schl-Teller [20],
the Woods-Saxon potential [21-23], the Kratzer-type potentials
[12,14,24-27], the Rosen-Morse-type potentials [28,29], the Manning-Rosen
potential [30-33], generalized Morse potential [34] and other multiparameter
exponential-type potentials [35]. Various methods are used to obtain the
exact solutions of the wave equations for this type of exponential
potentials. These methods include the supersymmetric (SUSY) and shape
invariant method [19,36], the variational [37], the path integral approach
[31], the standard methods [32,33], the asymptotic iteration method (AIM)
[38], the exact quantization rule (EQR) [13,39,40], the hypervirial
perturbation [41], the shifted $1/N$ expansion (SE) [42] and the modified
shifted $1/N$ expansion (MSE) [43], series method [44], smooth
transformation [45], the algebraic approach [46], the perturbative treatment
[47,48] and the Nikiforov-Uvarov (NU) method [16,17,20--26,49-51] and
others. The NU method [51] is based on solving the second-order linear
differential equation by reducing to a generalized equation of
hypergeometric type. It has been used to solve the Schr\"{o}dinger
[14,16,20,22,48,49], Dirac [17,28,34,50], Klein-Gordon [21,24,25,50] wave
equations for such kinds of exponential potentials.

The NU method has shown its power in calculating the exact energy levels of
all bound states for some solvable quantum systems. Motivated by the
considerable interest in exponential-type potentials [12-35], we attempt to
study the quantum properties of another exponential-type potential proposed
by Manning and Rosen (MR) [29-33]%
\begin{equation}
V(r)=\frac{\hbar ^{2}}{2\mu b^{2}}\left( \frac{\alpha (\alpha -1)e^{-2r/b}}{%
\left( 1-e^{-r/b}\right) ^{2}}-\frac{Ae^{-r/b}}{1-e^{-r/b}}\right) ,
\end{equation}%
where $A$ and $\alpha $ are two-dimensionless parameters but the screening
parameter $b$ has dimension of length and corresponds to the potential range
[33]. This potential is used as a methematical model in the description of
diatomic molecular vibrations [52,53] and it constitutes a convenient model
for other physical situations. Figure 1 plots the Manning-Rosen potential
(1) versus $r$ for various screening distances $b=0.025,$ $0.050,$ and $%
0.100 $ considering the cases (a) $\alpha =0.75$ and (b) $\alpha =1.50.$ It
is known that for this potential the Schr\"{o}dinger equation can be solved
exactly for $s$-wave (i.e., $l=0$) [32]. Unfortunately, for an arbitrary $l$%
-states ($l\neq 0),$ in which the Schr\"{o}dinger equation does not admit an
exact analytic solution. In such a case, the Schr\"{o}dinger equation is
solved numerically [54] or approximately using approximation schemes
[18,50,55,56,57]. Some authors used the approximation scheme proposed by
Greene and Aldrich [18] to study analytically the $l\neq 0$ bound states or
scattering states of the Schr\"{o}dinger or even relativistic wave equations
for MR potential [13,21]. We calculate and find its $l\neq 0$ bound state
energy spectrum and normalized wave functions [29-33]. The potential (1) may
be further put in the following simple form%
\begin{equation}
V(r)=-\frac{Ce^{-r/b}+De^{-2r/b}}{\left( 1-e^{-r/b}\right) ^{2}},\text{ }C=A,%
\text{ }D=-A-\alpha \text{(}\alpha -1)\text{,}
\end{equation}%
It is also used in several branches of physics for their bound states and
scattering properties. Its spectra have already been calculated via Schr\"{o}%
dinger formulation [30]. In our analysis, we find that the potential (1)
remains invariant by mapping $\alpha \rightarrow 1-\alpha .$ Further, it has
a relative minimum value $V(r_{0})=-\frac{A^{2}}{4\kappa b^{2}\alpha (\alpha
-1)}$ at $r_{0}=b\ln \left[ 1+\frac{2\alpha (\alpha -1)}{A}\right] $ for $%
A/2+\alpha (\alpha -1)>0$ which provides $2\alpha >1+\sqrt{1-2A}$ as a
result of the first derivative $\left. \frac{dV}{dr}\right\vert
_{r=r_{0}}=0. $ For the case $\alpha =0.75,$ we have the criteria imposed on
the value of $A$ is $A>\alpha /2=3/8.$ For example, in $\hbar =\mu =1,$ the
minimum of the potential is $V(r_{0})=-\alpha /16b^{2}(\alpha -1).$ The
second derivative which determines the force constants at $r=r_{0}$ is given
by%
\begin{equation}
\left. \frac{d^{2}V}{dr^{2}}\right\vert _{r=r_{0}}=\frac{A^{2}\left[
A+2\alpha (\alpha -1)\right] ^{2}}{8b^{4}\alpha ^{3}(\alpha -1)^{3}}.
\end{equation}%
The purpose of this paper is to investigate the $l$-state solution of the
Schr\"{o}dinger-MR problem within the Nikiforov-Uvarov method to generate
accurate energy spectrum. The solution is mainly depends on replacing the
orbital centrifugal term of singularity $\sim 1/r^{2}$ [17] with
Greene-Aldrich approximation scheme. consisting of the exponential form
[16]. Figure 2 shows the behaviour of the singular term $r^{-2}$ and various
approximation schemes recently used in Refs. [18,34,55,56].

The paper is organized as follows: In Section II we present the shortcuts of
the NU method. In Section III, we derive $l\neq 0$ bound state
eigensolutions (energy spectrum and wave functions) of the MR potential by
means of the NU method. In Section IV, we give numerical calculations for
various diatomic molecules. Section V, is devoted to for two special cases,
namely, $l=0$ and the Hulth\'{e}n potential. The concluding remarks are
given in Section VI.

\section{\noindent Method}

The Nikiforov-Uvarov (NU) method is based on solving the hypergeometric type
second order differential equation [51]. Employing an appropriate coordinate
transformation $z=z(r),$ we may rewrite the Schr\"{o}dinger equation in the
following form:%
\begin{equation}
\psi _{n}^{\prime \prime }(z)+\frac{\widetilde{\tau }(z)}{\sigma (z)}\psi
_{n}^{\prime }(z)+\frac{\widetilde{\sigma }(z)}{\sigma ^{2}(z)}\psi
_{n}(z)=0,
\end{equation}%
where $\sigma (z)$ and $\widetilde{\sigma }(z)$ are the polynomials with at
most of second-degree, and $\widetilde{\tau }(s)$ is a first-degree
polynomial. Further, using $\psi _{n}(z)=\phi _{n}(z)y_{n}(z),$ Eq. (4)
reduces into an equation of the following hypergeometric type:%
\begin{equation}
\sigma (z)y_{n}^{\prime \prime }(z)+\tau (z)y_{n}^{\prime }(z)+\lambda
y_{n}(z)=0,
\end{equation}%
where $\tau (z)=\widetilde{\tau }(z)+2\pi (z)$ (its derivative must be
negative) and $\lambda $ is a constant given in the form%
\begin{equation}
\lambda =\lambda _{n}=-n\tau ^{\prime }(z)-\frac{n\left( n-1\right) }{2}%
\sigma ^{\prime \prime }(z),\text{\ \ \ }n=0,1,2,...
\end{equation}%
It is worthwhile to note that $\lambda $ or $\lambda _{n}$ are obtained from
a particular solution of the form $y(z)=y_{n}(z)$ which is a polynomial of
degree $n.$ Further, $\ y_{n}(z)$ is the hypergeometric-type function whose
polynomial solutions are given by Rodrigues relation%
\begin{equation}
y_{n}(z)=\frac{B_{n}}{\rho (z)}\frac{d^{n}}{dz^{n}}\left[ \sigma ^{n}(z)\rho
(z)\right] ,
\end{equation}%
where $B_{n}$ is the normalization constant and the weight function $\rho
(z) $ must satisfy the condition [51]%
\begin{equation}
w^{\prime }(z)-\left( \frac{\tau (z)}{\sigma (z)}\right) w(z)=0,\text{ }%
w(z)=\sigma (z)\rho (z).
\end{equation}%
In order to determine the weight function given in Eq. (8), we must obtain
the following polynomial:%
\begin{equation}
\pi (z)=\frac{\sigma ^{\prime }(z)-\widetilde{\tau }(z)}{2}\pm \sqrt{\left( 
\frac{\sigma ^{\prime }(z)-\widetilde{\tau }(z)}{2}\right) ^{2}-\widetilde{%
\sigma }(z)+k\sigma (z)}.
\end{equation}%
In principle, the expression under the square root sign in Eq. (9) can be
arranged as the square of a polynomial. This is possible only if its
discriminant is zero. In this case, an equation for $k$ is obtained. After
solving this equation, the obtained values of $k$ are included in the NU
method and here there is a relationship between $\lambda $ and $k$ by $%
k=\lambda -\pi ^{\prime }(z).$ After this point an appropriate $\phi _{n}(z)$
can be calculated as the solution of the differential equation:%
\begin{equation}
\phi ^{\prime }(z)-\left( \frac{\pi (z)}{\sigma (z)}\right) \phi (z)=0.
\end{equation}

\section{Bound-state solutions for arbitrary $l$-states}

To study any quantum physical system characterized by the empirical
potential given in Eq. (1), we solve the original $\mathrm{SE}$ which is
given in the well known textbooks [1,2]

\begin{equation}
\left( \frac{p^{2}}{2m}+V(r)\right) \psi (\mathbf{r,}\theta ,\phi )=E\psi (%
\mathbf{r,}\theta ,\phi ),
\end{equation}%
where the potential $V(r)$ is taken as the MR form in (1). Using the
separation method with the wavefunction $\psi (\mathbf{r,}\theta ,\phi
)=r^{-1}R(r)Y_{lm}(\theta ,\phi ),$ we obtain the following radial Schr\"{o}%
dinger eqauation as%
\begin{equation}
\frac{d^{2}R_{nl}(r)}{dr^{2}}+\left\{ \frac{2\mu E_{nl}}{\hbar ^{2}}-\frac{1%
}{b^{2}}\left[ \frac{\alpha (\alpha -1)e^{-2r/b}}{\left( 1-e^{-r/b}\right)
^{2}}-\frac{Ae^{-r/b}}{1-e^{-r/b}}\right] -\frac{l(l+1)}{r^{2}}\right\}
R_{nl}(r)=0,
\end{equation}%
Since the Schr\"{o}dinger equation with above MR effective potential has no
analytical solution for $l\neq 0$ states$,$ an approximation to the
centrifugal term has to be made. The good approximation for the too singular
kinetic energy term $l(l+1)r^{-2}$ in the centrifugal barrier is taken as
[18,33]%
\begin{equation}
\frac{1}{r^{2}}\approx \frac{1}{b^{2}}\frac{e^{-r/b}}{\left(
1-e^{-r/b}\right) ^{2}},
\end{equation}%
in a short potential range. To solve it by the present method, we need to
recast Eq. (12) with Eq. (13) into the form of Eq. (4) by making change of
the variables $r\rightarrow z$ through the mapping function $r=f(z)$ and
energy transformation:%
\begin{equation}
z=e^{-r/b},\text{ }\varepsilon =\sqrt{-\frac{2\mu b^{2}E_{nl}}{\hbar ^{2}}},%
\text{ }E_{nl}<0,
\end{equation}%
to obtain the following hypergeometric equation:%
\[
\frac{d^{2}R(z)}{dz^{2}}+\frac{(1-z)}{z(1-z)}\frac{dR(z)}{dz} 
\]%
\begin{equation}
+\frac{1}{\left[ z(1-z)\right] ^{2}}\left\{ -\varepsilon ^{2}+\left[
A+2\varepsilon ^{2}-l(l+1)\right] z-\left[ A+\varepsilon ^{2}+\alpha (\alpha
-1)\right] z^{2}\right\} R(z)=0.
\end{equation}%
It is noted that the bound state (real) solutions of the last equation
demands that%
\begin{equation}
z=\left\{ 
\begin{array}{ccc}
0, & \text{when} & r\rightarrow \infty , \\ 
1, & \text{when} & r\rightarrow 0,%
\end{array}%
\right.
\end{equation}%
and thus provide the finite radial wave functions $R_{nl}(z)\rightarrow 0.$
To apply the hypergeometric method (NU), it is necessary to compare Eq. (15)
with Eq. (4). Subsequently, the following value for the parameters in Eq.
(4) are obtained as 
\begin{equation}
\widetilde{\tau }(z)=1-z,\text{\ }\sigma (z)=z-z^{2},\text{\ }\widetilde{%
\sigma }(z)=-\left[ A+\varepsilon ^{2}+\alpha (\alpha -1)\right] z^{2}+\left[
A+2\varepsilon ^{2}-l(l+1)\right] z-\varepsilon ^{2}.
\end{equation}%
If one inserts these values of parameters into Eq. (9), with $\sigma
^{\prime }(z)=1-2z,$ the following linear function is achieved%
\begin{equation}
\pi (z)=-\frac{z}{2}\pm \frac{1}{2}\sqrt{a_{1}z^{2}+a_{2}z+a_{3}},
\end{equation}%
where $a_{1}=1+4\left[ A+\varepsilon ^{2}+\alpha (\alpha -1)-k\right] ,$ $%
a_{2}=4\left\{ k-\left[ A+2\varepsilon ^{2}-l(l+1)\right] \right\} $ and $%
a_{3}=4\varepsilon ^{2}.$ According to this method the expression in the
square root has to be set equal to zero, that is, $\Delta
=a_{1}z^{2}+a_{2}z+a_{3}=0.$ Thus the constant $k$ can be determined as%
\begin{equation}
k=A-l(l+1)\pm a\varepsilon ,\text{ \ }a=\sqrt{(1-2\alpha )^{2}+4l(l+1)}.
\end{equation}%
In view of that, we can find four possible functions for $\pi (z)$ as%
\begin{equation}
\pi (z)=-\frac{z}{2}\pm \left\{ 
\begin{array}{c}
\varepsilon -\left( \varepsilon -\frac{a}{2}\right) z,\text{ \ \ \ for \ \ }%
k=A-l(l+1)+a\varepsilon , \\ 
\varepsilon -\left( \varepsilon +\frac{a}{2}\right) z;\text{ \ \ \ for \ \ }%
k=A-l(l+1)-a\varepsilon .%
\end{array}%
\right.
\end{equation}%
We must select%
\begin{equation}
\text{\ }k=A-l(l+1)-a\varepsilon ,\text{ }\pi (z)=-\frac{z}{2}+\varepsilon
-\left( \varepsilon +\frac{a}{2}\right) z,
\end{equation}%
in order to obtain the polynomial, $\tau (z)=\widetilde{\tau }(z)+2\pi (z)$
having negative derivative as%
\begin{equation}
\tau (z)=1+2\varepsilon -\left( 2+2\varepsilon +a\right) z,\text{ }\tau
^{\prime }(z)=-(2+2\varepsilon +a).
\end{equation}%
We can also write the values of $\lambda =k+\pi ^{\prime }(z)$ and $\lambda
_{n}=-n\tau ^{\prime }(z)-\frac{n\left( n-1\right) }{2}\sigma ^{\prime
\prime }(z),$\ $n=0,1,2,...$ as%
\begin{equation}
\lambda =A-l(l+1)-(1+a)\left[ \frac{1}{2}+\varepsilon \right] ,
\end{equation}%
\begin{equation}
\lambda _{n}=n(1+n+a+2\varepsilon ),\text{ }n=0,1,2,...
\end{equation}%
respectively. Letting $\lambda =\lambda _{n}$ and solving the resulting
equation for $\varepsilon $ leads to the energy equation%
\begin{equation}
\varepsilon =\frac{(n+1)^{2}+l(l+1)+(2n+1)\Lambda -A}{2(n+1+\Lambda )},\text{
}\Lambda =\frac{-1+a}{2},
\end{equation}%
from which we obtain the discrete energy spectrum formula:%
\begin{equation}
E_{nl}=-\frac{\hbar ^{2}}{2\mu b^{2}}\left[ \frac{(n+1)^{2}+l(l+1)+(2n+1)%
\Lambda -A}{2(n+1+\Lambda )}\right] ^{2},\text{ \ }0\leq n,l<\infty
\end{equation}%
where $n$ denotes the radial quantum number. It is found that $\Lambda $
remains invariant by mapping $\alpha \rightarrow 1-\alpha ,$ so do the bound
state energies $E_{nl}.$ An important quantity of interest for the MR
potential is the critical coupling constant $A_{c},$ which is that value of $%
A$ for which the binding energy of the level in question becomes zero.
Furthermore, from Eq. (26), we have (in atomic units $\hbar =\mu =Z=e=1),$%
\begin{equation}
A_{c}=(n+1+\Lambda )^{2}-\Lambda (\Lambda +1)+l(l+1).
\end{equation}

Next, we turn to the radial wave function calculations. We use $\sigma (z)$
and $\pi (z)$ in Eq (17) and Eq. (21) to obtain%
\begin{equation}
\phi (z)=z^{\varepsilon }(1-z)^{\Lambda +1},
\end{equation}%
and weight function%
\begin{equation}
\rho (z)=z^{2\varepsilon }(1-z)^{2\Lambda +1},
\end{equation}%
\begin{equation}
y_{nl}(z)=C_{n}z^{-2\varepsilon }(1-z)^{-(2\Lambda +1)}\frac{d^{n}}{dz^{n}}%
\left[ z^{n+2\varepsilon }(1-z)^{n+2\Lambda +1}\right] .
\end{equation}%
The functions $\ y_{nl}(z)$, up to a numerical factor, are in the form of\
Jacobi polynomials, i.e., $\ y_{nl}(z)\simeq P_{n}^{(2\varepsilon ,2\Lambda
+1)}(1-2z),$ and physically holds in the interval $(0\leq r<\infty $ $%
\rightarrow $ $0\leq z\leq 1)$ [58]. Therefore, the radial part of the wave
functions can be found by substituting Eq. (28) and Eq. (30) into $%
R_{nl}(z)=\phi (z)y_{nl}(z)$ as%
\begin{equation}
R_{nl}(z)=N_{nl}z^{\varepsilon }(1-z)^{1+\Lambda }P_{n}^{(2\varepsilon
,2\Lambda +1)}(1-2z),
\end{equation}%
where $\varepsilon $ and $\Lambda $ are given in Eqs. (14) and (19) and $%
N_{nl}$ is a normalization constant. This equation satisfies the
requirements; $R_{nl}(z)=0$ as $z=0$ $(r\rightarrow \infty )$ and $%
R_{nl}(z)=0$ as $z=1$ $(r=0).$ Therefore, the wave functions, $R_{nl}(z)$ in
Eq. (31) is valid physically in the closed interval $z\in \lbrack 0,1]$ or $%
r\in (0,\infty ).$ Further, the wave functions satisfy the normalization
condition:%
\begin{equation}
\int\limits_{0}^{\infty }\left\vert R_{nl}(r)\right\vert
^{2}dr=1=b\int\limits_{0}^{1}z^{-1}\left\vert R_{nl}(z)\right\vert ^{2}dz,
\end{equation}%
where $N_{nl}$ can be determined via%
\begin{equation}
1=bN_{nl}^{2}\int\limits_{0}^{1}z^{2\varepsilon -1}(1-z)^{2\Lambda +2}\left[
P_{n}^{(2\varepsilon ,2\Lambda +1)}(1-2z)\right] ^{2}dz.
\end{equation}%
The Jacobi polynomials, $P_{n}^{(\rho ,\nu )}(\xi ),$ can be explicitly
written in two different ways [59,60]::%
\begin{equation}
P_{n}^{(\rho ,\nu )}(\xi )=2^{-n}\sum\limits_{p=0}^{n}(-1)^{n-p}\binom{%
n+\rho }{p}\binom{n+\nu }{n-p}\left( 1-\xi \right) ^{n-p}\left( 1+\xi
\right) ^{p},
\end{equation}

\begin{equation}
P_{n}^{(\rho ,\nu )}(\xi )=\frac{\Gamma (n+\rho +1)}{n!\Gamma (n+\rho +\nu
+1)}\sum\limits_{r=0}^{n}\binom{n}{r}\frac{\Gamma (n+\rho +\nu +r+1)}{\Gamma
(r+\rho +1)}\left( \frac{\xi -1}{2}\right) ^{r},
\end{equation}%
where $\binom{n}{r}=\frac{n!}{r!(n-r)!}=\frac{\Gamma (n+1)}{\Gamma
(r+1)\Gamma (n-r+1)}.$ After using Eqs. (34) and (35), we obtain the
explicit expressions for $P_{n}^{(2\varepsilon ,2\Lambda +1)}(1-2z):$%
\[
P_{n}^{(2\varepsilon ,2\Lambda +1)}(1-2z)=(-1)^{n}\Gamma (n+2\varepsilon
+1)\Gamma (n+2\Lambda +2) 
\]

\begin{equation}
\times \sum\limits_{p=0}^{n}\frac{(-1)^{p}}{p!(n-p)!\Gamma (p+2\Lambda
+2)\Gamma (n+2\varepsilon -p+1)}z^{n-p}(1-z)^{p},
\end{equation}

\begin{equation}
P_{n}^{(2\varepsilon ,2\Lambda +1)}(1-2z)=\frac{\Gamma (n+2\varepsilon +1)}{%
\Gamma (n+2\varepsilon +2\Lambda +2)}\sum\limits_{r=0}^{n}\frac{%
(-1)^{r}\Gamma (n+2\varepsilon +2\Lambda +r+2)}{r!(n-r)!\Gamma (2\varepsilon
+r+1)}z^{r}.
\end{equation}%
Inserting Eqs. (36) and (37) into Eq. (33), one obtains%
\[
1=bN_{nl}^{2}(-1)^{n}\frac{\Gamma (n+2\Lambda +2)\Gamma (n+2\varepsilon
+1)^{2}}{\Gamma (n+2\varepsilon +2\Lambda +2)} 
\]%
\begin{equation}
\times \sum\limits_{p,r=0}^{n}\frac{(-1)^{p+r}\Gamma (n+2\varepsilon
+2\Lambda +r+2)}{p!r!(n-p)!(n-r)!\Gamma (p+2\Lambda +2)\Gamma
(n+2\varepsilon -p+1)\Gamma (2\varepsilon +r+1)}I_{nl}(p,r),
\end{equation}%
where%
\begin{equation}
I_{nl}(p,r)=\int\limits_{0}^{1}z^{n+2\varepsilon +r-p-1}(1-z)^{p+2\Lambda
+2}dz.
\end{equation}%
Using the following integral representation of the hypergeometric function
[59.60]%
\[
_{2}F_{1}(\alpha _{0},\beta _{0}:\gamma _{0};1)\frac{\Gamma (\alpha
_{0})\Gamma (\gamma _{0}-\alpha _{0})}{\Gamma (\gamma _{0})}%
=\int\limits_{0}^{1}z^{\alpha _{0}-1}(1-z)^{\gamma _{0}-\alpha
_{0}-1}(1-z)^{-\beta _{0}}dz, 
\]

\begin{equation}
\func{Re}(\gamma _{0})>\func{Re}(\alpha _{0})>0,
\end{equation}%
which gives%
\begin{equation}
_{2}F_{1}(\alpha _{0},\beta _{0}:\alpha _{0}+1;1)/\alpha
_{0}=\int\limits_{0}^{1}z^{\alpha _{0}-1}(1-z)^{-\beta _{0}}dz,
\end{equation}%
where%
\[
_{2}F_{1}(\alpha _{0},\beta _{0}:\gamma _{0};1)=\frac{\Gamma (\gamma
_{0})\Gamma (\gamma _{0}-\alpha _{0}-\beta _{0})}{\Gamma (\gamma _{0}-\alpha
_{0})\Gamma (\gamma _{0}-\beta _{0})}, 
\]%
\begin{equation}
(\func{Re}(\gamma _{0}-\alpha _{0}-\beta _{0})>0,\text{ }\func{Re}(\gamma
_{0})>\func{Re}(\beta _{0})>0).
\end{equation}%
For the present case, with the aid of Eq. (40), when $\alpha
_{0}=n+2\varepsilon +r-p,$ $\beta _{0}=-p-2\Lambda -2,$ and $\gamma
_{0}=\alpha _{0}+1$ are substituted into Eq. (41)$,$ we obtain%
\begin{equation}
I_{nl}(p,r)=\frac{_{2}F_{1}(\alpha _{0},\beta _{0}:\gamma _{0};1)}{\alpha
_{0}}=\frac{\Gamma (n+2\varepsilon +r-p+1)\Gamma (p+2\Lambda +3)}{%
(n+2\varepsilon +r-p)\Gamma (n+2\varepsilon +r+2\Lambda +3)}.
\end{equation}%
Finally, we obtain%
\[
1=bN_{nl}^{2}(-1)^{n}\frac{\Gamma (n+2\Lambda +2)\Gamma (n+2\varepsilon
+1)^{2}}{\Gamma (n+2\varepsilon +2\Lambda +2)} 
\]%
\begin{equation}
\times \sum\limits_{p,r=0}^{n}\frac{(-1)^{p+r}\Gamma (n+2\varepsilon
+r-p+1)(p+2\Lambda +2)}{p!r!(n-p)!(n-r)!\Gamma (n+2\varepsilon -p+1)\Gamma
(2\varepsilon +r+1)(n+2\varepsilon +r+2\Lambda +2)},
\end{equation}%
which gives%
\begin{equation}
N_{nl}=\frac{1}{\sqrt{s(n)}},
\end{equation}%
where%
\[
s(n)=b(-1)^{n}\frac{\Gamma (n+2\Lambda +2)\Gamma (n+2\varepsilon +1)^{2}}{%
\Gamma (n+2\varepsilon +2\Lambda +2)} 
\]%
\begin{equation}
\times \sum\limits_{p,r=0}^{n}\frac{(-1)^{p+r}\Gamma (n+2\varepsilon
+r-p+1)(p+2\Lambda +2)}{p!r!(n-p)!(n-r)!\Gamma (n+2\varepsilon -p+1)\Gamma
(2\varepsilon +r+1)(n+2\varepsilon +r+2\Lambda +2)}.
\end{equation}

\section{Numerical Results}

To show the accuracy of our results, we calculate the energy eigenvalues for
various $n$ and $l$ quantum numbers with two different values of the
parameters $\alpha .$ Its shown in Table 1, the present approximately
numerical results are not in a good agreement when long potential range
(small values of parameter $b$). The energy eigenvalues for short potential
range (large values of parameter $b$) are in agreement with the other
authors. The energy spectra for various diatomic molecules like $HCl,CH,LiH$
and $CO$ are presented in Tables 2 and 3. These results are relevant to
atomic physics [61-64], molecular physics [65,66] and chemical physics
[67,68], etc.

\section{Discussions}

In this work, we have utilized the hypergeometric method and solved the
radial $\mathrm{SE}$ for the M-R model potential with the angular momentum $%
l\neq 0$ states$.$ We have derived the binding energy spectra in Eq. (26)
and their corresponding wave functions in Eq. (31).

Let us study special cases. We have shown that for $\alpha =0$ $(1)$, the
present solution reduces to the one of the Hulth\'{e}n potential [16,19,57]:%
\begin{equation}
V^{(H)}(r)=-V_{0}\frac{e^{-\delta r}}{1-e^{-\delta r}},\text{ }%
V_{0}=Ze^{2}\delta ,\text{ }\delta =b^{-1}
\end{equation}%
where $Ze^{2}$ is the potential strength parameter and $\delta $ is the
screening parameter and $b$ is the range of potential. We note also that it
is possible to recover the Yukawa potential by letting $b\rightarrow \infty $
and $V_{0}=Ze^{2}/b.$ If the potential is used for atoms, the $Z$ is
identified with the atomic number. This can be achieved by setting $\Lambda
=l,$ hence, the energy for $l\neq 0$ states%
\begin{equation}
E_{nl}=-\frac{\left[ A-(n+l+1)^{2}\right] ^{2}\hbar ^{2}}{8\mu
b^{2}(n+l+1)^{2}},\text{ \ }0\leq n,l<\infty .
\end{equation}%
and for $s$-wave ($l=0)$ states%
\begin{equation}
E_{n}=-\frac{\left[ A-(n+1)^{2}\right] ^{2}\hbar ^{2}}{8\mu b^{2}(n+1)^{2}},%
\text{ \ }0\leq n<\infty
\end{equation}%
Essentially, these results coincide with those obtained by the Feynman
integral method [31,56] and the standard way [32,33], respectively.
Furthermore, if taking $b=1/\delta $ and identifying $\frac{A\hbar ^{2}}{%
2\mu b^{2}}$ as $Ze^{2}\delta ,$ we are able to obtain%
\begin{equation}
E_{nl}=-\frac{\mu \left( Ze^{2}\right) ^{2}}{2\hbar ^{2}}\left[ \frac{1}{%
n+l+1}-\frac{\hbar ^{2}\delta }{2Ze^{2}\mu }(n+l+1)\right] ^{2},
\end{equation}%
which coincides with those of Refs. [16,19]. Further, we have (in atomic
units $\hbar =\mu =Z=e=1)$%
\begin{equation}
E_{nl}=-\frac{1}{2}\left[ \frac{1}{n+l+1}-\frac{(n+l+1)}{2}\delta \right]
^{2},
\end{equation}%
which coincides with Refs. [16,33].

The corresponding radial wave functions are expressed as%
\begin{equation}
R_{nl}(r)=N_{nl}e^{-\delta \varepsilon r}(1-e^{-\delta
r})^{l+1}P_{n}^{(2\varepsilon ,2l+1)}(1-2e^{-\delta r}),
\end{equation}%
where%
\begin{equation}
\varepsilon =\frac{\mu Ze^{2}}{\hbar ^{2}\delta }\left[ \frac{1}{n+l+1}-%
\frac{\hbar ^{2}\delta }{2Ze^{2}\mu }(n+l+1)\right] ,\text{ }0\leq
n,l<\infty ,
\end{equation}%
which coincides for the ground state with that given in Eq. (6) by G\"{o}n%
\"{u}l \textit{et al.} [18]. In addition, for $\delta r\ll 1$ (i.e., $r/b\ll
1),$ the Hulth\'{e}n potential turns to become a Coulomb potential: $%
V(r)=-Ze^{2}/r$ with energy levels and wave functions:%
\[
E_{nl}=-\frac{\varepsilon _{0}}{(n+l+1)^{2}},\text{ }n=0,1,2,.. 
\]

\begin{equation}
.\varepsilon _{0}=\frac{Z^{2}\hbar ^{2}}{2\mu a_{0}^{2}},\text{ }a_{0}=\frac{%
\hbar ^{2}}{\mu e^{2}}
\end{equation}%
where $\varepsilon _{0}=13.6$ $eV$ and $a_{0}$ is Bohr radius for the
Hydrogen atom. The wave functions are%
\[
R_{nl}=N_{nl}\exp \left[ -\frac{\mu Ze^{2}}{\hbar ^{2}}\frac{r}{\left(
n+l+1\right) }\right] r^{l+1}P_{n}^{\left( \frac{2\mu Ze^{2}}{\hbar
^{2}\delta (n+l+1)},2l+1\right) }(1+2\delta r) 
\]%
which coincide with Refs. [3,16,22].

\section{Conclusions and Outlook}

In this work approximately analytical bound states for the $l$-wave Schr\"{o}%
dinger equationwith the MR potential have been presented by making a proper
approximation to the too singular orbital centrifugal term $\sim r^{-2}.$
The normalized radial wave functions of $l$-wave bound states associated
with the MR potential are obtained. The approach enables one to find the $l$%
-dependent solutions and the corresponding energy eigenvalues for different
screening parameters of the MR potential.

We have shown that \ for $\alpha =0,1,$ the present solution reduces to the
one of the Hulth\'{e}n potential. We note that it is possible to recover the
Yukawa potential by letting $b\rightarrow \infty $ and $V_{0}=Ze^{2}/b.$ The
Hulth\'{e}n potential behaves like the Coulomb potential near the origin
(i.e., $r\rightarrow 0$) $V_{C}(r)=-Ze^{2}/r$ but decreases exponentially in
the asymptotic region when $r\gg 0,$ so its capacity for bound states is
smaller than the Coulomb potential [16]. Obviously, the results are in good
agreement with those obtained by other methods for short potential range,
small $\alpha $ and $l.$ We have also studied two special cases for $l=0,$ $%
l\neq 0$ and Hulth\'{e}n potential. The results we have ended up show that
the NU method constitute a reliable alternative way in solving the
exponential potentials. We have also found that the criteria for the choice
of parameter $A$ requires that $A$ satisfies the inequality $\sqrt{1-2A}%
<2\alpha -1.$ This means that for real bound state solutions $A$ should be
chosen properly in our numerical calculations.

A slight difference in the approximations of the numerical energy spectrum
of Schr\"{o}dinger-MR problem is found in Refs. [55,56] and present work
since the approximation schemes are different by a small shift $\delta
^{2}/12.$ In our recent work [17], we have found that the physical
quantities like the energy spectrum are critically dependent on the behavior
of the system near the singularity ($r=0$). That is why, for example, the
energy spectrum depends strongly on the angular momentum $l$, which results
from the $r^{-2}$ singularity of the orbital term, even for high excited
states. It is found that the $r^{-2\text{ }}$ orbital term is too singular,
then the validity of all such approximations is limited only to very few of
the lowest energy states. In this case, to extend accuracy to higher energy
states one may attempt to utilize the full advantage of the unique features
of Schr\"{o}dinger equation. Therefore, it is more fruitful to perform the
analytic approximation of the less singularity $r^{-1}$ rather than the too
singular term $r^{-2}$ which makes it possible to extend the validity of the
results to higher excitation levels giving better analytic approximation for
a wider energy spectrum [69].

\acknowledgments Work partially supported by the Scientific and
Technological Research Council of Turkey (T\"{U}B\.{I}TAK).

\newpage

{\normalsize 
}

\bigskip

\bigskip

\bigskip \newpage

\bigskip {\normalsize 
}

\bigskip

\bigskip \baselineskip= 2\baselineskip
\bigskip

\FRAME{ftbpFO}{0.0277in}{0.0277in}{0pt}{\Qct{Variation of MR potential as
function of separation distance $r$ taking various values for the screening
parameter $b$ when (a) $\protect\alpha =0.75$ and (b) $\protect\alpha =1.50.$%
}}{}{Figure 1}{}\FRAME{ftbpFO}{0.0277in}{0.0277in}{0pt}{\Qct{A plot of the
variation of the singular orbital term $1/r^{2}$ (dotted-solid line) with
the approximations of (a) Ref. 34 (dash line), the conventional
Greene-Aldrich of Ref. 18 (dash-dot line) and improved [55,56] (solid line)
replacing the term $1/r^{2}$ with respect to $r$ where $\protect\delta =0.1$ 
$fm^{-1},$ and (b) the improved approximation [55] with various shifting
constants.}}{}{Figure 2}{} 
\begin{table}[tbp]
\caption{Energies (in atomic units) of different $n$ and $l$ states and for $%
\protect\alpha =0.75$ and $\protect\alpha =1.5,$ $A=2b.$}%
\begin{tabular}{llllllll}
&  & $\alpha =0.75$ &  &  & $\alpha =1.5$ &  &  \\ 
states & $1/b$ & Present & QD [33] & LSl [54] & Present & QD [33] & LS [54]
\\ 
\tableline$2p$ & $0.025$ & $-0.1205793$ & $-0.1205793$ & $-0.1205271$ & $%
-0.0900228$ & $-0.0900229$ & $-0.0899708$ \\ 
& $0.050$ & $-0.1084228$ & $-0.1084228$ & $-0.1082151$ & $-0.0802472$ & $%
-0.0802472$ & $-0.0800400$ \\ 
& $0.075$ & $-0.0969120$ & $-0.0969120$ & $-0.0964469$ & $-0.0710332$ & $%
-0.0710332$ & $-0.0705701$ \\ 
& $0.100$ & $-0.0860740$ &  &  & $-0.0577157$ &  &  \\ 
$3p$ & $0.025$ & $-0.0459296$ & $-0.0459297$ & $-0.0458779$ & $-0.0369650$ & 
$-0.0369651$ & $-0.0369134$ \\ 
& $0.050$ & $-0.0352672$ & $-0.0352672$ & $-0.0350633$ & $-0.0274719$ & $%
-0.0274719$ & $-0.0272696$ \\ 
& $0.075$ & $-0.0260109$ & $-0.0260110$ & $-0.0255654$ & $-0.0193850$ & $%
-0.0193850$ & $-0.0189474$ \\ 
& $0.100$ & $-0.0181609$ &  &  & $-0.0127043$ &  &  \\ 
$3d$ & $0.025$ & $-0.0449299$ & $-0.0449299$ & $-0.0447743$ & $-0.0396344$ & 
$-0.0396345$ & $-0.0394789$ \\ 
& $0.050$ & $-0.0343082$ & $-0.0343082$ & $-0.0336930$ & $-0.0300629$ & $%
-0.0300629$ & $-0.0294496$ \\ 
& $0.075$ & $-0.0251168$ & $-0.0251168$ & $-0.0237621$ & $-0.0218120$ & $%
-0.0218121$ & $-0.0204663$ \\ 
$4p$ & $0.025$ & $-0.0208608$ & $-0.0208608$ & $-0.0208097$ & $-0.0172249$ & 
$-0.0172249$ & $-0.0171740$ \\ 
& $0.050$ & $-0.0119291$ & $-0.0119292$ & $-0.0117365$ & $-0.0091019$ & $%
-0.0091019$ & $-0.0089134$ \\ 
& $0.075$ & $-0.0054773$ & $-0.0054773$ & $-0.0050945$ & $-0.0035478$ & $%
-0.0035478$ & $-0.0031884$ \\ 
$4d$ & $0.025$ & $-0.0204555$ & $-0.0204555$ & $-0.0203017$ & $-0.0183649$ & 
$-0.0183649$ & $-0.0182115$ \\ 
& $0.050$ & $-0.0115741$ & $-0.0115742$ & $-0.0109904$ & $-0.0100947$ & $%
-0.0100947$ & $-0.0095167$ \\ 
& $0.075$ & $-0.0052047$ & $-0.0052047$ & $-0.0040331$ & $-0.0042808$ & $%
-0.0042808$ & $-0.0031399$ \\ 
$4f$ & $0.025$ & $-0.0202886$ & $-0.0202887$ & $-0.0199797$ & $-0.0189222$ & 
$-0.0189223$ & $-0.0186137$ \\ 
& $0.050$ & $-0.0114283$ & $-0.0114284$ & $-0.0102393$ & $-0.0105852$ & $%
-0.0105852$ & $-0.0094015$ \\ 
& $0.075$ & $-0.0050935$ & $-0.0050935$ & $-0.0026443$ & $-0.0046527$ & $%
-0.0046527$ & $-0.0022307$ \\ 
$5p$ & $0.025$ & $-0.0098576$ & $-0.0098576$ & $-0.0098079$ & $-0.0081308$ & 
$-0.0081308$ & $-0.0080816$ \\ 
$5d$ & $0.025$ & $-0.0096637$ & $-0.0096637$ & $-0.0095141$ & $-0.0086902$ & 
$-0.0086902$ & $-0.0085415$ \\ 
$5f$ & $0.025$ & $-0.0095837$ & $-0.0095837$ & $-0.0092825$ & $-0.0089622$ & 
$-0.0089622$ & $-0.0086619$ \\ 
$5g$ & $0.025$ & $-0.0095398$ & $-0.0095398$ & $-0.0090330$ & $-0.0091210$ & 
$-0.0091210$ & $-0.0086150$ \\ 
$6p$ & $0.025$ & $-0.0044051$ & $-0.0044051$ & $-0.0043583$ & $-0.0035334$ & 
$-0.0035334$ & $-0.0034876$ \\ 
$6d$ & $0.025$ & $-0.0043061$ & $-0.0043061$ & $-0.0041650$ & $-0.0038209$ & 
$-0.0038209$ & $-0.0036813$ \\ 
$6f$ & $0.025$ & $-0.0042652$ & $-0.0042652$ & $-0.0039803$ & $-0.0039606$ & 
$-0.0039606$ & $-0.0036774$ \\ 
$6g$ & $0.025$ & $-0.0042428$ & $-0.0042428$ & $-0.0037611$ & $-0.0040422$ & 
$-0.0040422$ & $-0.0035623$%
\end{tabular}%
\end{table}

\begin{table}[tbp]
\caption{Energy spectrum of $HCl$ and $CH$ (in $eV$) for different states
where $\hbar c=1973.29$ $eV$ $A^{\circ },$ $\protect\mu _{HCl}=0.9801045$ $%
amu,$ $\protect\mu _{CH}=0.929931$ $amu$ and $A=2b.$}%
\begin{tabular}{llllllll}
states & $1/b$\tablenotetext[1]{$b$ is in $pm$.}\tablenotemark[1] & $HCl/$ $%
\alpha =0,1$ & $\alpha =0.75$ & $\alpha =1.5$ & $CH/$ $\alpha =0,1$ & $%
\alpha =0.75$ & $\alpha =1.5$ \\ 
\tableline$2p$ & $0.025$ & $-4.81152646$ & $-5.14278553$ & $-3.83953094$ & $%
-5.07112758$ & $-5.42025940$ & $-4.04668901$ \\ 
& $0.050$ & $-4.31837832$ & $-4.62430290$ & $-3.42259525$ & $-4.55137212$ & $%
-4.87380256$ & $-3.60725796$ \\ 
& $0.075$ & $-3.85188684$ & $-4.13335980$ & $-3.02961216$ & $-4.05971155$ & $%
-4.35637111$ & $-3.19307186$ \\ 
& $0.100$ & $-3.41205201$ & $-3.66996049$ & $-2.46161213$ & $-3.59614587$ & $%
-3.86796955$ & $-2.59442595$ \\ 
$3p$ & $0.025$ & $-1.86633700$ & $-1.95892730$ & $-1.57658128$ & $%
-1.96703335 $ & $-2.06461927$ & $-1.66164415$ \\ 
& $0.050$ & $-1.42316902$ & $-1.50416901$ & $-1.17169439$ & $-1.49995469$ & $%
-1.58532495$ & $-1.23491200$ \\ 
& $0.075$ & $-1.03998066$ & $-1.10938179$ & $-0.82678285$ & $-1.09609178$ & $%
-1.16923738$ & $-0.87139110$ \\ 
& $0.100$ & $-0.71676763$ & $-0.77457419$ & $-0.54184665$ & $-0.75544012$ & $%
-0.81636557$ & $-0.57108145$ \\ 
$3d$ & $0.025$ & $-1.86633700$ & $-1.91628944$ & $-1.69043293$ & $%
-1.96703335 $ & $-2.01968093$ & $-1.78163855$ \\ 
& $0.050$ & $-1.42316902$ & $-1.46326703$ & $-1.28220223$ & $-1.49995469$ & $%
-1.54221615$ & $-1.35138217$ \\ 
& $0.075$ & $-1.03998066$ & $-1.07124785$ & $-0.93029598$ & $-1.09609178$ & $%
-1.12904596$ & $-0.98048917$ \\ 
& $0.100$ & $-0.71676763$ & $-0.74022762$ & $-0.63472271$ & $-0.75544012$ & $%
-0.78016587$ & $-0.66896854$ \\ 
$4p$ & $0.025$ & $-0.85301300$ & $-0.88972668$ & $-0.73465318$ & $%
-0.89903647 $ & $-0.93773100$ & $-0.77429066$ \\ 
& $0.050$ & $-0.47981981$ & $-0.50878387$ & $-0.38820195$ & $-0.50570801$ & $%
-0.53623480$ & $-0.40914700$ \\ 
& $0.075$ & $-0.21325325$ & $-0.23361041$ & $-0.15131598$ & $-0.22475912$ & $%
-0.24621462$ & $-0.15948008$ \\ 
$4d$ & $0.025$ & $-0.85301300$ & $-0.87244037$ & $-0.78327492$ & $%
-0.89903647 $ & $-0.91951202$ & $-0.82553574$ \\ 
& $0.050$ & $-0.47981981$ & $-0.49364289$ & $-0.43054552$ & $-0.50570801$ & $%
-0.52027690$ & $-0.45377517$ \\ 
& $0.075$ & $-0.21325325$ & $-0.22198384$ & $-0.18257890$ & $-0.22475912$ & $%
-0.23396076$ & $-0.19242977$ \\ 
$4f$ & $0.025$ & $-0.85301300$ & $-0.86532198$ & $-0.80704413$ & $%
-0.89903647 $ & $-0.91200956$ & $-0.85058739$ \\ 
& $0.050$ & $-0.47981981$ & $-0.48742442$ & $-0.45146566$ & $-0.50570801$ & $%
-0.51372292$ & $-0.47582404$ \\ 
& $0.075$ & $-0.21325325$ & $-0.21724109$ & $-0.19844068$ & $-0.22475912$ & $%
-0.22896211$ & $-0.20914735$ \\ 
$5p$ & $0.025$ & $-0.40318193$ & $-0.42043305$ & $-0.34678391$ & $%
-0.42493521 $ & $-0.44311709$ & $-0.36549429$ \\ 
$5d$ & $0.025$ & $-0.40318193$ & $-0.41216309$ & $-0.37064268$ & $%
-0.42493521 $ & $-0.43440094$ & $-0.39064034$ \\ 
$5f$ & $0.025$ & $-0.40318193$ & $-0.40875104$ & $-0.38224366$ & $%
-0.42493521 $ & $-0.43080479$ & $-0.40286723$ \\ 
$5g$ & $0.025$ & $-0.40318193$ & $-0.40687867$ & $-0.38901658$ & $%
-0.42493521 $ & $-0.42883140$ & $-0.41000558$ \\ 
$6p$ & $0.025$ & $-0.17919244$ & $-0.18788038$ & $-0.15070181$ & $%
-0.18886059 $ & $-0.19801728$ & $-0.15883277$ \\ 
$6d$ & $0.025$ & $-0.17919244$ & $-0.18365796$ & $-0.16296387$ & $%
-0.18886059 $ & $-0.19356705$ & $-0.17175642$ \\ 
$6f$ & $0.025$ & $-0.17919244$ & $-0.18191355$ & $-0.16892216$ & $%
-0.18886059 $ & $-0.19172852$ & $-0.17803620$ \\ 
$6g$ & $0.025$ & $-0.17919244$ & $-0.18095818$ & $-0.17240246$ & $%
-0.18886059 $ & $-0.19072160$ & $-0.18170426$%
\end{tabular}%
\end{table}

\begin{table}[tbp]
\caption{Energy spectrum of $LiH$ and $CO$ (in $eV$) for different states
where $\hbar c=1973.29$ $eV$ $A^{\circ },$ $\protect\mu _{LiH}=0.8801221$ $%
amu,$ $\protect\mu _{CO}=6.8606719$ $amu$ and $A=2b.$}%
\begin{tabular}{llllllll}
states & $1/b$\tablenotemark[1]\tablenotetext[1]{$b$ is in $pm$.} & $LiH/$ $%
\alpha =0,1$ & $\alpha =0.75$ & $\alpha =1.5$ & $CO/$ $\alpha =0,1$ & $%
\alpha =0.75$ & $\alpha =1.5$ \\ 
\tableline$2p$ & $0.025$ & $-5.35811876$ & $-5.72700906$ & $-4.27570397$ & $%
-1.374733789$ & $-0.734690030$ & $-0.548509185$ \\ 
& $0.050$ & $-4.80894870$ & $-5.14962650$ & $-3.81140413$ & $-1.233833096$ & 
$-0.660620439$ & $-0.488946426$ \\ 
& $0.075$ & $-4.28946350$ & $-4.60291196$ & $-3.37377792$ & $-1.100548657$ & 
$-0.590485101$ & $-0.432805497$ \\ 
& $0.100$ & $-3.79966317$ & $-4.08687021$ & $-2.74125274$ & $-0.974880471$ & 
$-0.524284624$ & $-0.351661930$ \\ 
$3p$ & $0.025$ & $-2.07835401$ & $-2.18146262$ & $-1.75568186$ & $%
-0.533243776$ & $-0.279849188$ & $-0.225227854$ \\ 
& $0.050$ & $-1.58484188$ & $-1.67504351$ & $-1.30479958$ & $-0.406623254$ & 
$-0.214883153$ & $-0.167386368$ \\ 
& $0.075$ & $-1.15812308$ & $-1.23540823$ & $-0.92070588$ & $-0.297139912$ & 
$-0.158484490$ & $-0.118112862$ \\ 
& $0.100$ & $-0.79819287$ & $-0.86256629$ & $-0.60340076$ & $-0.204792531$ & 
$-0.110654417$ & $-0.077407337$ \\ 
$3d$ & $0.025$ & $-2.07835401$ & $-2.13398108$ & $-1.88246712$ & $%
-0.533243776$ & $-0.273758013$ & $-0.241492516$ \\ 
& $0.050$ & $-1.58484188$ & $-1.62949505$ & $-1.42786117$ & $-0.406623254$ & 
$-0.209039964$ & $-0.183173338$ \\ 
& $0.075$ & $-1.15812308$ & $-1.19294225$ & $-1.03597816$ & $-0.299139912$ & 
$-0.153036736$ & $-0.132900580$ \\ 
& $0.100$ & $-0.79819287$ & $-0.82431793$ & $-0.70682759$ & $-0.204792531$ & 
$-0.105747722$ & $-0.090675460$ \\ 
$4p$ & $0.025$ & $-0.94991579$ & $-0.99080017$ & $-0.81811023$ & $%
-0.243720118$ & $-0.127104916$ & $-0.104951366$ \\ 
& $0.050$ & $-0.53432763$ & $-0.56658202$ & $-0.43230193$ & $-0.137092566$ & 
$-0.072684041$ & $-0.055457903$ \\ 
& $0.075$ & $-0.23747895$ & $-0.26014869$ & $-0.16850556$ & $-0.060930029$ & 
$-0.033373205$ & $-0.021616756$ \\ 
$4d$ & $0.025$ & $-0.94991579$ & $-0.97155012$ & $-0.87225543$ & $%
-0.243720118$ & $-0.124635422$ & $-0.111897390$ \\ 
& $0.050$ & $-0.53432763$ & $-0.54972102$ & $-0.47945575$ & $-0.137092566$ & 
$-0.070521025$ & $-0.061507037$ \\ 
& $0.075$ & $-0.23747895$ & $-0.24720134$ & $-0.20331998$ & $-0.060930029$ & 
$-0.031712252$ & $-0.026082927$ \\ 
$4f$ & $0.025$ & $-0.94991579$ & $-0.96362308$ & $-0.89872483$ & $%
-0.243720118$ & $-0.123618500$ & $-0.115293020$ \\ 
& $0.050$ & $-0.53432763$ & $-0.54279613$ & $-0.50275243$ & $-0.137092566$ & 
$-0.069632666$ & $-0.064495655$ \\ 
& $0.075$ & $-0.23747895$ & $-0.24191980$ & $-0.22098366$ & $-0.060930029$ & 
$-0.031034710$ & $-0.028348915$ \\ 
$5p$ & $0.025$ & $-0.44898364$ & $-0.46819450$ & $-0.38617877$ & $%
-0.115195837$ & $-0.060062386$ & $-0.049540988$ \\ 
$5d$ & $0.025$ & $-0.44898364$ & $-0.45898506$ & $-0.41274791$ & $%
-0.115195837$ & $-0.058880953$ & $-0.052949414$ \\ 
$5f$ & $0.025$ & $-0.44898364$ & $-0.45518540$ & $-0.42566677$ & $%
-0.115195837$ & $-0.058393512$ & $-0.054606711$ \\ 
$5g$ & $0.025$ & $-0.44898364$ & $-0.45310033$ & $-0.43320910$ & $%
-0.115195837$ & $-0.058126029$ & $-0.055574280$ \\ 
$6p$ & $0.025$ & $-0.19954881$ & $-0.20922370$ & $-0.16782162$ & $%
-0.051198285$ & $-0.026840287$ & $-0.021529017$ \\ 
$6d$ & $0.025$ & $-0.19954881$ & $-0.20452162$ & $-0.18147666$ & $%
-0.051198285$ & $-0.026237080$ & $-0.023280755$ \\ 
$6f$ & $0.025$ & $-0.19954881$ & $-0.20257904$ & $-0.18811182$ & $%
-0.051198285$ & $-0.025987876$ & $-0.024131947$ \\ 
$6g$ & $0.025$ & $-0.19954881$ & $-0.20151514$ & $-0.19198748$ & $%
-0.051198285$ & $-0.025851393$ & $-0.024629136$%
\end{tabular}%
\end{table}

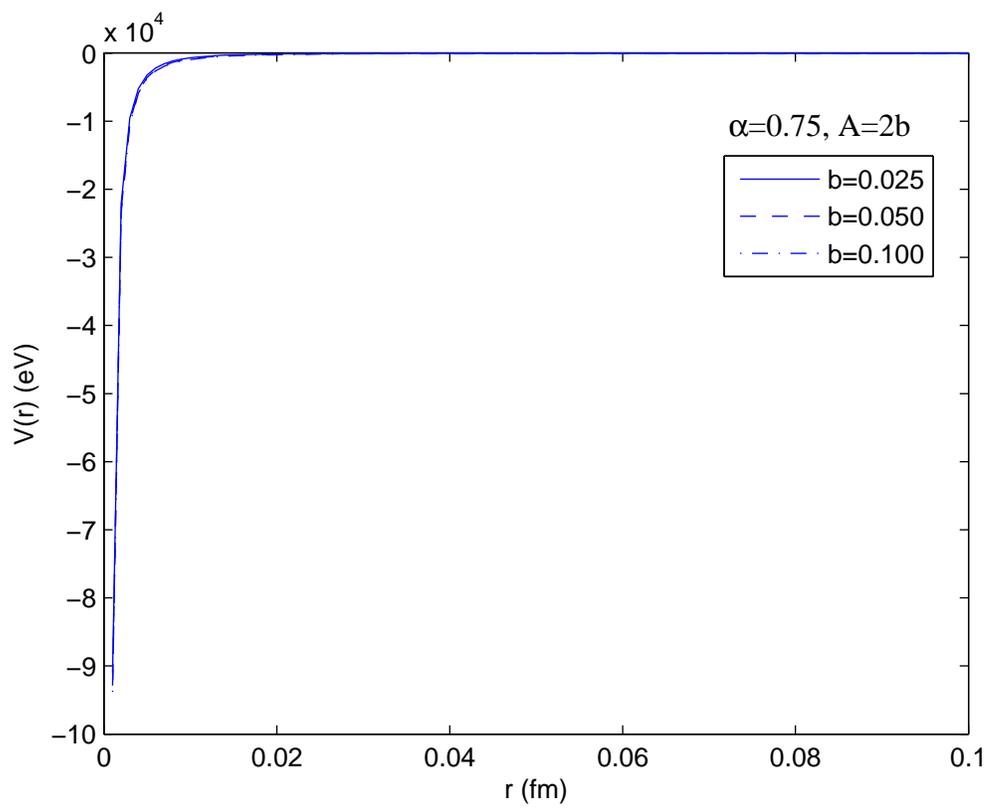
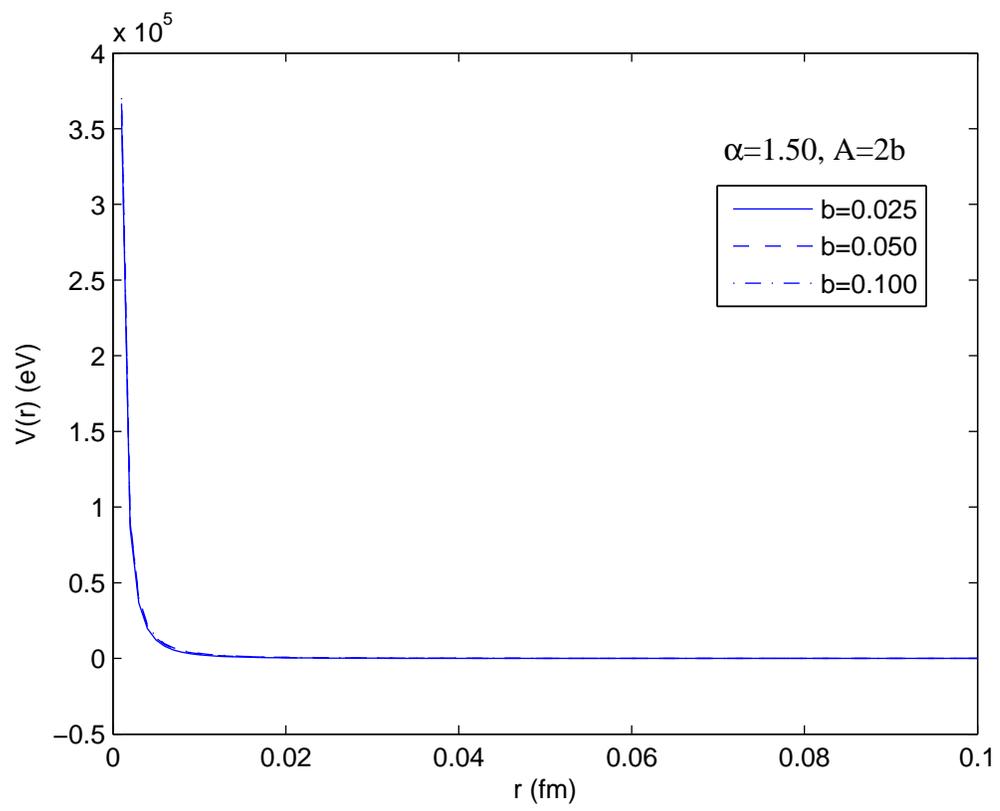
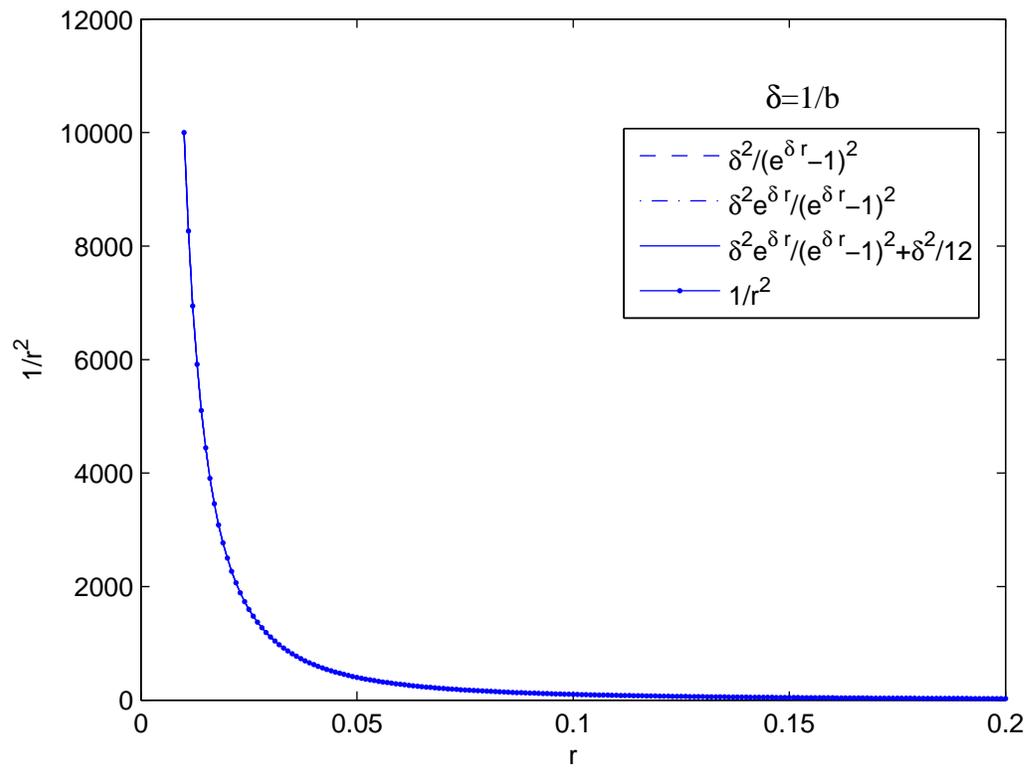
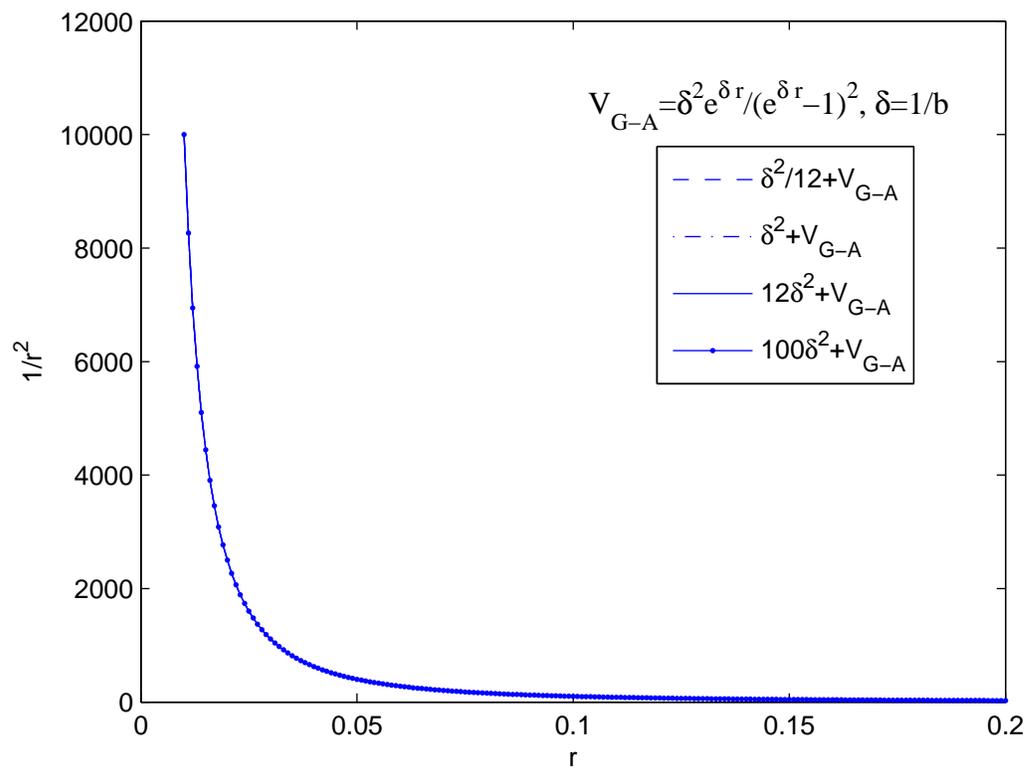
\end{document}